\documentclass[fleqn,usenatbib]{mnras}

\usepackage{newtxtext,newtxmath}

\usepackage[T1]{fontenc}

\DeclareRobustCommand{\VAN}[3]{#2}
\let\VANthebibliography\thebibliography
\def\thebibliography{\DeclareRobustCommand{\VAN}[3]{##3}\VANthebibliography}

\usepackage{ulem}
\usepackage{graphicx}	% Including figure files
\usepackage{amsmath}	% Advanced maths commands
\usepackage{booktabs}   %tables
\usepackage{xcolor}     % colors

\title[Dynamical model of PDS~70]%{A dynamical model of the PDS~70 system}
{Planet migration, resonant locking and accretion streams in PDS~70: Comparing models and data}
\author[C. Toci al.]{
Claudia Toci,$^{1,2}$\thanks{E-mail: claudia.toci@inaf.it}
Giuseppe Lodato,$^{2}$
Valentin Christiaens,$^{3}$
Davide Fedele,$^{4}$
\newauthor
Christophe Pinte,$^{3}$
Daniel J. Price,$^{3}$
Leonardo Testi$^{5}$
\\
$^{1}$INAF OA Brera, via Brera 28, 20121, Milano MI, Italy \\
$^{2}$Dipartimento di Fisica, Università degli Studi di Milano, Via Celoria 16, 20133 Milano MI, Italy\\
$^{3}$School of Physics and Astronomy, Monash University, VIC 3800, Australia \\
$^{4}$INAF OA Arcetri, Largo Enrico Fermi 5, 50125 Firenze FI \\
$^{5}$European Southern Observatory, Karl-Schwarzschild-Strasse 2, 85748 Garching bei Munchen, Germany
}

\date{Accepted 2020 September 19. Received 2020 September 18; in original form 2020 August 10}

\pubyear{2020}

\begin{document}
\label{firstpage}
\pagerange{\pageref{firstpage}--\pageref{lastpage}}
\maketitle

% Abstract of the paper
\begin{abstract}
The disc surrounding PDS 70, with two directly imaged embedded giant planets, is an ideal laboratory to study planet-disc interaction. We present three-dimensional smoothed particle hydrodynamics simulations of the system. In our simulations, planets, which are free to migrate and accrete mass, end up in a locked resonant configuration that is dynamically stable. We show that features observed at infrared (scattered light) and millimetre (thermal continuum) wavelengths are naturally explained by the accretion stream onto the outer planet, without requiring a circumplanetary disc around planet c.
We post-processed our near-infrared synthetic images in order to account for observational biases known to affect high-contrast images. Our successful reproduction of the observations indicates that planet-disc dynamical interactions alone are sufficient to explain the observations of PDS 70.
\end{abstract}

% Select between one and six entries from the list of approved keywords.
% Don't make up new ones.

%%%%%%%%%%%%%%%%%%%%%%%%%%%%%%%%%%%%%%%%%%%%%%%%%%

%%%%%%%%%%%%%%%%% BODY OF PAPER %%%%%%%%%%%%%%%%%%

\section{Introduction}\label{sec:intro}

Thanks to the exceptional capabilities of new observing facilities such as the Atacama Large Millimeter Array
(ALMA) and the Spectro-Polarimetric High-contrast Exoplanet
REsearch (SPHERE) instrument at the Very Large Telescope, high angular resolution observations of protoplanetary discs are now available. These high-resolution observations have dramatically changed our view of disc evolution \citep{Testi2015}: small-scale substructures such as rings, gaps, cavities and spiral arms appear to be common (e.g.  \citealt{Benisty2017,Andrews2018}). Interpreting these as signatures of planet-disc interactions (e.g., \citealt{Dipierro2015,Pinilla2012,Fedele2018,Dipierro2018,Pinte2018,Zhang2019,Lodato2019}) suggests that giant planets may be already fully formed at an age of only 1 Myr or less. 

This explanation is however not unique --- other mechanisms have been invoked as the cause of the observed features, such as self-induced reconnection in magnetised disc-winds  \citep{Suriano2017}, dust trapping \citep{Gonzalez2015} and vortices \citep{Barge2017}.
Planet signatures may appear in other disc diagnostics, such as in the kinematics of gas as traced by CO lines \citep{Pinte2018,Teague2018,Pinte2019,Casassus2019}. In some cases planets detected with kinematical methods lie within observed dust gaps, strengthening the idea of planet-carved gaps. 

PDS 70 is a $\sim$5 Myr old \citep{Muller2018} K7-type T-Tauri star located at a distance of 113.4 $\pm$ 0.5~pc \citep{Gaia2018}, within the Scorpius-Centaurus association \citep{Pecaut2016}. This young star is surrounded by a protoplanetary disc, first inferred from infrared excess in the  spectral energy distribution \citep{Gregorio2002} and lately directly imaged with different techniques in the near infrared and sub-mm wavelengths \citep{Hashimoto2012,Dong2012,Hashimoto2015,Long2018,Muller2018,Keppler2019,Christiaens2019,Isella2019,Mesa2019}. The disc has a wide gap (from $\sim 20-50$ au) clear of dust in the near infrared observations, that becomes larger ($\sim 74$ au) when observed at sub-mm wavelengths both in the dust \citep{Keppler2018} and in the CO emission \citep{Keppler2019}. This suggests that dust grains of different sizes are spatially segregated by the radial pressure gradient of the disc, an hypothesis tested numerically in \citet{Bae2019}. 

Two planets, PDS 70b and c, have been directly imaged in the disc \citep{Keppler2018,Muller2018,Christiaens2019,Haffert2019} at 195 and 234 milliarcsecond (mas),
corresponding to deprojected distances of $\sim 20$ and $35$~au, respectively. The latter suggest that the planets might be in mean-motion resonance ($\sim 2:1$, as pointed out by \citealt{Haffert2019} and \citealt{Bae2019}). Estimates for the mass of the planets range from 4-17 Jupiter Masses (hereafter M$_{\rm J}$) for PDS 70b \citep{Muller2018,Christiaens2019b,Haffert2019} and 2-10~M$_{\rm J}$ for PDS 70c \citep{Haffert2019,Bae2019,Isella2019}. Based on the excess emission of $H\alpha$  at the location of the protoplanet candidates, \citet{Haffert2019} estimated a mass accretion rate of $\sim 2~\times~10^{-8}$~M$_{\rm J}$~yr$^{-1}$ for the inner planet PDS 70b and $\sim 10^{-8}$~M$_{\rm J}$~yr$^{-1}$ for the outer planet PDS 70c, while the stellar accretion rate is $\sim 5.5~\times~10^{-8}$~M$_{\rm J}$ yr$^{-1}$ \citep{Haffert2019}. However, these measurements are uncertain.  

The two planets can be responsible for the dynamical clearing of the inner disc and for the dust trap \citep{Pinilla2012}. Observations in the sub-mm \citep{Isella2019} and IR \citep{Christiaens2019b} wavelengths suggest the presence of a circumplanetary disc surrounding each planet, a possibility that has been theoretically predicted (see e.g., \citealt{Lubow1999}) and tested using numerical simulations (e.g., \citealt{Ayliffe1999}, \citealt{Szulagy2017} for a numerical study on the formation mechanisms of circumplanetary discs around giant planets or \citealt{Szulagy2018} for numerical estimates of the detection range of these structures using ALMA). Moreover, $^{12}$CO-integrated intensity maps \citep{Keppler2019} shows evidence of an inner decreasing of the flux density, located at $\sim$ 0.2 arcsec, most likely carved by PDS 70b, and an outer decreasing at $\sim$ 0.6 arcsec, given by the dust being optically thick.

Interestingly, images of PDS 70 show sub-mm emission near (but not coincident with the expected location of the planet) PDS 70b \citep{Isella2019}, and several features in scattered light close to the theoretical position of PDS 70c, present in 2.11/2.15 $\mu$m scattered light images, close to a broader and brighter feature \citep{Keppler2018, Muller2018}. The nature of these features is still uncertain.

\citet{Bae2019} performed two dimensional simulations of this source. They showed that the presence of the inner planet alone (PDS 70b) fails to reproduce the observed peak in the radial profile of the sub-mm observation of \citep{Keppler2019}, as well as the spatially unresolved continuum emission close to PDS 70b of \citet{Isella2019}.
They initialised their simulations with two grown giant planets placed close to the 2:1 mean motion resonance, testing different values for the mass of PDS 70c while fixing the mass of PDS 70b to 5~M$_{\rm J}$, in order to find the best configuration to reproduce the observations. They focussed on dust trapping, size segregation and filtration. \citet{Bae2019} also discussed whether or not the observed unresolved continuum flux in the vicinity of the two planets \citep{Isella2019} could be explained with circumplanetary (CPD) discs, with their study providing an order of magnitude estimate of their dust mass, $\sim 2.2 \times 10^{-5}$ and $4.5\times 10^{-5}$ Earth masses respectively. 

In this work we present new 3D smoothed particle hydrodynamics (SPH) simulations of the PDS 70 disc assuming two embedded giant planets. We model both dust and gas, aiming to understand the physical origin of features observed in mm and scattered light images. 
In contrast to \citet{Bae2019}, we start our simulations with lower mass planets and let them grow and migrate to eventually reach a mean motion resonance. 
This condition better represents a real dynamical scenario for planets embedded in the disc. 
In order to analyse how sensitive the final configuration is to the initial position and mass of the planets, we present the results of three simulations with different initial conditions.

The paper is organised as follows: in Sec.~\ref{sec:Methods} we describe the simulation initial conditions and radiative transfer models used to obtain synthetic observations. A comparison between our models and observational data and the dynamics of the planets are shown in Sec.~\ref{sec:Results}. We discuss the implication of our results in Sec.~\ref{sec:Discussion} and give conclusions in Sec.~\ref{sec:conclusions}. 

\section{Methods}\label{sec:Methods}
We perform a set of three dust and gas hydrodynamics simulations in 3D using \textsc{phantom}, a smoothed particle hydrodynamics (SPH) code \citep{Price2018}.
SPH (see \citealt{Monaghan2005} or \citealt{Price2012} for reviews) is particularly suitable for the study of accretion flows and asymmetric interactions between planets and the disc, because there is no preferred geometry. Angular momentum is conserved to machine precision, allowing for accurate orbital dynamics. \textsc{phantom} has already been used widely for studies of dust and gas in discs (e.g. \citealt{Dipierro2015, Ragusa2017, Cuello2018, Mentiplay2019, Gerrit2019,Veronesi2019, Toci2020}). 
In this work, we assume a single grain size of $1.00$~mm, employing the \textit{one fluid} algorithm \citep{Laibe2014,Price2015,Ballabio2018} based on the terminal velocity approximation \citep{YG2005}. 
In all our simulations, we include the dust back-reaction and the gravity between the gas and dust discs and the planets.
\subsection{Simulation set} 
We present a small sample of three simulations where we fix the disc parameters and we vary the initial conditions of two embedded giant planets. 
The system is modelled with a central star of mass ${\rm M}_\star = 0.8~{\rm M}_\odot$ and the two giant planets embedded in a dust and gas disc. The planets are allowed to migrate and accrete mass.

In all the simulations our time unit is the
period of the outer planet (Planet c) at its initial position in Sim~2 and 3, $T_{\rm orb}$ ($T_{\rm orb}\sim$336~yr). All of our calculations have been evolved for at least 500 outer planet orbits, about $\sim$ 0.18~Myr.

\subsection{Gas and dust}\label{subsec:gas_dust}
We set up an initial surface density profile for the disc as in \citealt{LodatoPrice2010}, representing the disc with $10^6$ SPH particles. The value of the initial inner and outer radii are and $R_{\rm in}=5$~au in Sim~1 and $R_{\rm in}=2$~au in Sim~2 and 3. The outer radius is fixed in all the simulations, $R_{\rm out}=120$~au. The initial surface density profile for the gas is  
\begin{equation}\label{eq:sigma_in}
\Sigma(R)= \Sigma_0\left(\frac{R}{R_0}\right)^{-p}\exp{\left[ -(R/R_0)^{(2-q)}\right]}\left(1-\sqrt{\frac{R_{\rm{in}}}{R}}\right), 
\end{equation}
where $R_0=100$~au, the density normalisation $\Sigma_0=1.07$ g~cm$^{-2}$ is set to have an initial total gas mass of $\sim 10^{-2}~\rm{M}_\odot$ and we choose $p=1$, consistent with \citet{Bae2019}. 
We assume a locally isothermal equation of state with a sound speed profile
\begin{equation}
c_{\rm s} = c_{\rm{s,0}}\left(\frac{R}{R_{0}}\right)^{-q},
\end{equation}
where $q=0.5$ (again, following \citealt{Bae2019}). This assumption leads to the following aspect ratio for the gas 
\begin{equation}
\frac{H}{R} = \left(\frac{H_0}{R_0}\right) \left(\frac{R}{R_{0}}\right)^{(1/2-q)},
\end{equation}
where we set $H/R_0=0.05$ at $R=R_{\rm 0}$. In all our simulations we set the value of $\alpha_{\rm SPH}$ viscosity as in \citealt{LodatoPrice2010}, in order to have a value for the effective \citealt{SS1973}  viscosity of $\alpha_{\rm SS}=0.005$.
We initialise the dust with the same surface density profile as for the gas (Eq. \ref{eq:sigma_in}). The dust-to-gas ratio is initially
assumed constant for the whole disc extent and equal to 0.01. The initial dust mass is then $\rm{M}_{\rm dust}=10^{-4}~\rm{M}_\odot$. In all our dust simulation the chosen size of the dust is $a=1$~mm, and we assume an internal density of the dust of $\rho_{\rm d}=1$~g~cm$^{-2}$. 
For these values the initial midplane Stokes number, defined as $S_t = \rho_{\rm d} a / \Sigma_{\rm gas}$, is
smaller than unity. This implies that dust and the gas are initially coupled in the disc. This justifies the use of the \textit{one fluid} {\sc phantom} mode \citep{Laibe2014}.

% Init. cond. planets table
\begin{table}
    \centering
    \caption{Initial conditions for the three simulations, called sim~1, sim~2, sim~3. Initial conditions for the masses are in Jupiter mass (M$_{\rm J}$), the units for the initial planet separations are au while the sink radii $R_{\rm acc}$ are measured in fractions $f$ of the Hill radii (see sec \ref{sec:init_con_plan} for the definition of Hill radius).}\label{table:tab1}
    \begin{tabular}{*4c}
        \toprule
        & \multicolumn{3}{c}{Simulations} \\
        \cmidrule(lr){2-4}
               & Sim 1  & Sim 2  & Sim 3 \\    
%        Name   &        &         &         \\
        \midrule
        $M_{\rm b}$ ($M_{\rm J}$)     & 2.5 & 3   & 3   \\ 
        $R_{\rm b}$ (au)        & 20  & 15  & 20  \\
        $f_{\rm b}= R_{\rm acc,b}/R_{\rm{H}}$ & 1/8 & 1/4 & 1/4 \\
        $M_{\rm c}$ ($M_{\rm J}$)     & 0.5 & 0.5 & 0.5 \\
        $R_{\rm c}$ (au)        & 55  & 45  & 45  \\
        $f_{\rm c}= R_{\rm acc,c}/R_{\rm H}$ & 1/3 & 1/4 & 1/4 \\
        $R_{\rm c}/R_{\rm b}$      & 2.75  & 3 & 2.25 \\
        \bottomrule
    \end{tabular}
\end{table}

\subsection{Properties of the embedded planets}\label{sec:init_con_plan} 
The planets and the central star are included in the code as
sink particles, characterized by an initial mass, position and sink radius. In \textsc{phantom}, sinks are free to migrate and are able to
accrete gas and dust, interacting with dust and gas particle through gravitational forces (for an overview of sink particles in SPH see \citealt{Bate1995} and \citealt{Price2018}). 
All the material that enters the sink radius is considered as instantaneously accreted on the sink \footnote{Technically instant accretion occurs at 80\% of the sink radius by default, particles are accreted also at 80-100\% the sink radius in special occasions such as if they are bound or infalling}. 
We chose the sink radius of each planet to be a fraction $f$ of the Hill radius $R_H$, defined as
\begin{equation}\label{eq:Hill}
R_{\rm H}=\left(\frac{1}{3}\frac{M_{\rm p}}{M_\star}\right)^{1/3}R_{\rm p},    
\end{equation}
where $M_{\rm p}$ and $R_{\rm p}$ are the planet mass and radial distance from the star. The values of $f$ for the two planets are different in the simulations, and are listed in Table~\ref{table:tab1}.  The value of the sink radius may affect the planetary accretion rate and thus their final mass, so we decided to check the dependence of our results on this parameter.

The initial conditions for the mass and the position of the planets are significantly different from previous models \citep{Bae2019}. 
The fact that dust and gas can accrete onto the sinks represents a condition that mimics a real accretion scenario,
and the fact that planets are free to migrate due to planet-disc interaction, planet-planet interaction and viscous disc evolution allow us to set up our simulation without assuming that the planets are already close to the mean motion resonance as inferred from observations. Our aim is indeed to test if planets that are initially massive enough to open a gap in the gas surface density according to \citet{Crida2006} but are initially
smaller in masses and further out in position with respect to the observational results can evolve due to accretion and migration and obtain a final configuration with masses, position and motion resonances comparable with the observational results. 
The initial conditions for the planets can be found in Table \ref{table:tab1}. In all simulations the initial mass of the inner planet (M$_{\rm b}$) is heavier than the one of the outer planet (M$_{\rm c}$) and closer to the observed value, while the mass of the outer planet is far smaller than the observed value. This choice is motivated by our previous study on discs with two embedded giant planets \citep{Toci2020}: the initial mass of the outer planet has to be small enough to allow it to migrate inward, while the higher mass of the inner planet prevents this phenomenon. We also expect Planet c to accrete more mass than Planet b because of the larger gas and dust surface density in its neighbourhood.

\subsection{Radiative transfer}

To compare our simulations with observations, we post-processed our simulation using the Monte Carlo radiative transfer code \textsc{mcfost} \citep{Pinte2006,Pinte2009}. \textsc{mcfost} maps the dust and gas densities from the SPH particles to the radiative transfer grid, using a Voronoi tesselation to match each cell with an SPH particle, avoiding the need to interpolate.

The dust distribution in each cell is obtained by assuming dust grains smaller than 1\,$\mu$m follow the gas, and by linear interpolation in log of the grain size between 1\,$\mu$m and 1\,mm. 
We assume a power-law grain size dust distribution $dn/da \propto a^{-3.5}$ (integrating over the whole disc), spanning the value in a range from $a_{min}=0.03$~$\mu$m and $a_{max}=2500$~$\mu$m. Grains larger than 1\,mm are assumed to follow the distribution of the 1\,mm grains.
The dust optical properties are computed assuming
spherical and homogeneous dust grains (Mie theory), with chemical composition of 60$\%$ astronomical silicates and 15$\%$ amorphous carbons assumed, as the \textsc{diana} standard dust composition \citep{Woitke2016}, and a porosity of 10$\%$.
The total gas mass is directly taken from our SPH simulation and the total dust mass is derived from the gas to ratio of the SPH simulation, which is 100. The effective local dust to gas ratio for the various simulations is discussed in Sec. \ref{sec:Results}. 
We consider the emission of the central source as a black body with an effective temperature T$_{\rm e}$ = 4090 K and a radius of R$_\star$= 1.57 Solar radii, as derived from stellar evolution models for a 0.8\,M$_\odot$  at 5Myr \citep{Siess2000}.

In all our simulations, we fixed the values of the source distance $d=113$ pc \citep{Gaia2018}, disc inclination $i=180 - 49^{\circ}$ ( following \citet{Muller2018} to have an observed clockwise orbital motion of the planets $i$ > 90$^{\circ}$) and position angle PA=$159^{\circ}$. 
We use  $1.3\times 10^8$  photon packets in the Monte Carlo simulation to compute the temperature and specific intensities of our models. We then compute the source function by ray-tracing to produce the output images at the desired wavelengths.
We produced 855 $\mu$m continuum images and both 2.11$\mu$m and 2.25 $\mu$m scattered light images of the t = 500 T$_{\rm orb}$ ($\sim$ 0.18 Myr) snapshot for all three simulations. 
We also generated $^{12}$CO moment 0 map, to test if our models can reproduce the flux value and the observed depth of the cavity in CO of \citet{Keppler2019} observations. 
To compute the temperature and the synthetic line maps, following \citet{Pinte2018} and \citet{Pinte2019}, we assumed local thermal equilibrium (LTE), $T_{\rm gas} = T_{\rm dust}$ and a constant  $^{12}$CO abundance of $10^{-4}$. We added a local spatially unresolved turbulent velocity of 50 m s$^{-1}$ and we also considered photo-dissociation of molecules where the UV radiation is high. Precisely, as can be found in \citet{Pinte2018_2}, the condition for the photo-dissociation of the CO is $\log(\chi/n) > -6 $, where $n$ is the number density of hydrogen atoms  and  $\chi$ is the intensity of the UV radiation field. The intensity of the UV radiation field is evaluated as \citet{Woitke2009}:
\begin{equation}
\chi=\frac{1}{F^{{\rm Draine}}}\int_{91.2 nm}^{205 nm} u_\lambda d\lambda,
\end{equation}
where $F^{{\rm Draine}} =1.921 \times 10^{-12} m^{-1} s^{-1}$ and where $u_\lambda$ is the energy density of the radiation field computed by MCFOST in each point of the model.
Finally, we convolved the sub-mm images with a Gaussian beam to match the beam size of the observations (74 $\times$ 57 mas for the 855 $\mu$m image from \citealt{Keppler2019}. The IR synthetic images are further processed as explained in Section \ref{IRpostprocessing}.

\subsection{Post-processing of IR synthetic images} \label{IRpostprocessing}

The observation techniques and post-processing algorithms used to obtain high-contrast IR images of discs induce geometric biases to the incoming signals. In particular, angular differential imaging \citep[ADI;][]{Marois2006} is known to induce significant distortion to extended disc signals \citep[e.g.][]{Milli2012,Christiaens2019}. For a meaningful comparison of our synthetic IR image to observations of PDS 70, we have both reduced the 2018-02-24 SPHERE/IRDIS dataset presented in \citet[][ESO program 1100.C-0481]{Muller2018}, and post-processed our IR predictions in a similar way to the observations.

We calibrated the IRDIS data with a custom-made python pipeline which performs dark subtraction, flat-fielding, bad pixel correction, sky subtraction, star centering based on satellite spots, and gaussian fit of the unsaturated PSF in order to infer stellar flux and FWHM. The pipeline uses routines of the Vortex Image Processing package \citep[VIP;][]{GomezGonzalez2017}, an open-source compilation of python routines used in high-contrast imaging. We then processed the calibrated IRDIS cube using the median-ADI algorithm \citep{Marois2006} as in \citet{Muller2018}.

In order to mimic biases induced by high-contrast observations and image processing, we applied the following effects to the MCFOST images of PDS 70 predicted at 2.11 and 2.25 $\mu$m (i.e. at the central wavelength of the IRDIS K1 and K2 filters, respectively):
\begin{enumerate}
    \item Injection of the flux of protoplanets PDS 70 b and c at the location of the corresponding sink particles,
    using the contrast with respect to the star measured in \citet{Muller2018,Mesa2019};
    \item Convolution with the non-coronagraphic normalized point spread function observed in the IRDIS 2018-02-24 dataset, which has a FWHM of $\sim$63mas and $\sim$66mas in K1 and K2, respectively;
    \item Dampening by the radial transmission of SPHERE's Apodized Lyot Coronagraph \citep{Guerri2011};
    \item Resampling of the RT images to the same pixel scale as the observations (12.25mas/pixel) using a fourth-order Lanczos interpolation;
    \item Injection of shot noise at a similar level as in the observations;
    \item Creation of a synthetic ADI cube by rotating the RT images to match the parallactic angle coverage of the observations;
    \item Smearing of images by the expected field rotation in 96s-long exposures throughout the sequence;  
    \item Image processing using the median-ADI algorithm \citep{Marois2006};
    \item Averaging of the two median-ADI images obtained in the K1 and K2 filters.
\end{enumerate}
More details on steps (ii) to (vi) can be found in \citet{Calcino2020}, while the other steps were implemented specifically for this work. In particular, we added the effect of smearing given the long exposure time used in the IRDIS 2018-02-24 dataset. In order to estimate the amount of smearing to be applied in each image of the synthetic ADI cube, we used the initial, maximum and final parallactic rotation rates of the observed sequence and interpolated between these points using a second order polynomial.

\begin{figure}
\centering
 \includegraphics[width=0.85\columnwidth]{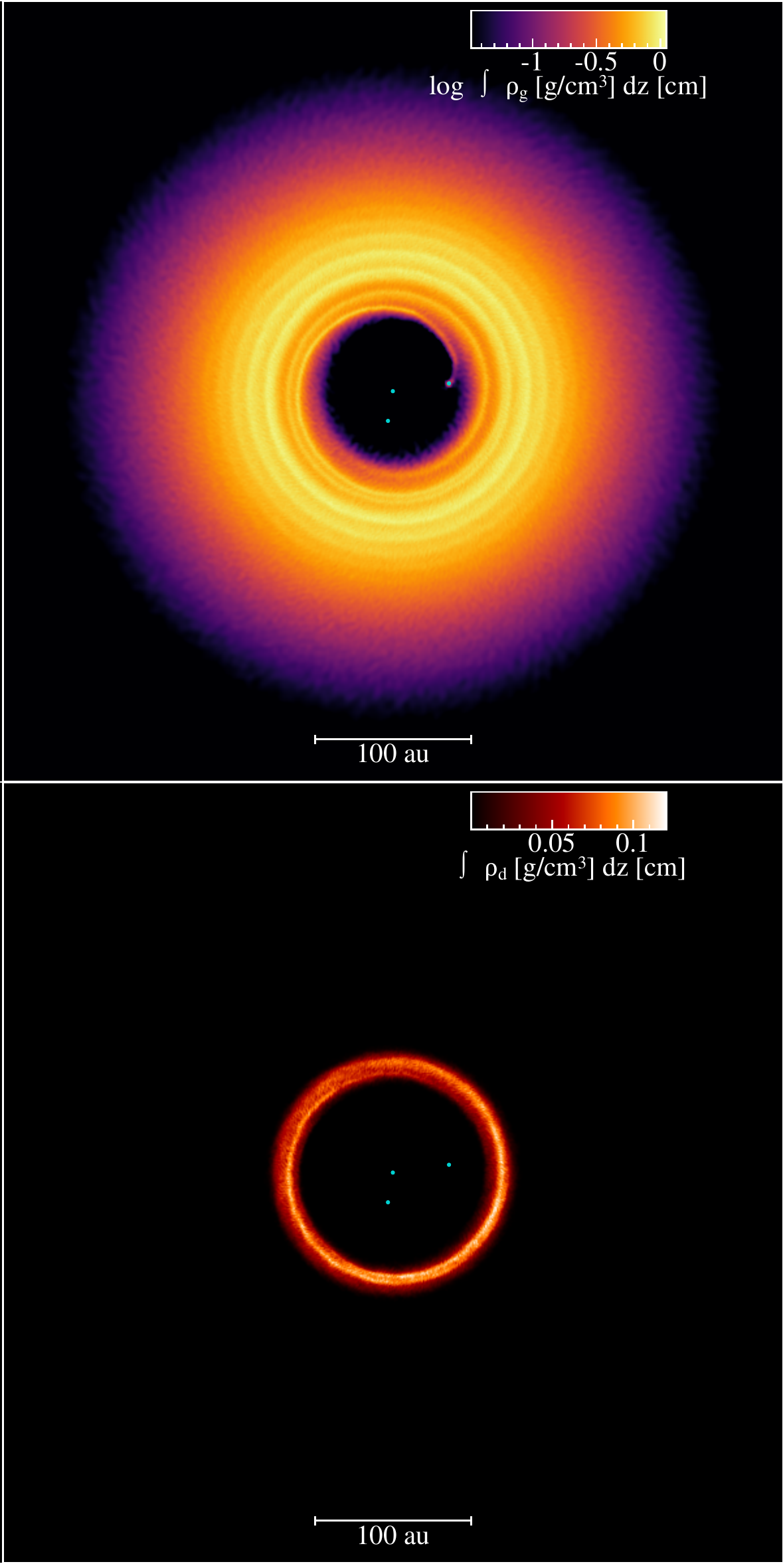}
 \caption{Top: logarithmic plot of the gas surface density map of the t~=~500~T$_{\rm orb}$ ($\sim$ 0.18 Myr) snapshot for Sim~3. The blue dots are the positions of the star and the planets. Bottom: linear plot of the 1 mm dust surface density map of the t = 500 T$_{\rm orb}$ ($\sim$ 0.18 Myr) snapshot for Sim~3. The blue dots are the positions of the star and the planets.}
 \label{fig:fig_surf_gas}
\end{figure}

\begin{figure}
\centering
 \includegraphics[width=0.85\columnwidth]{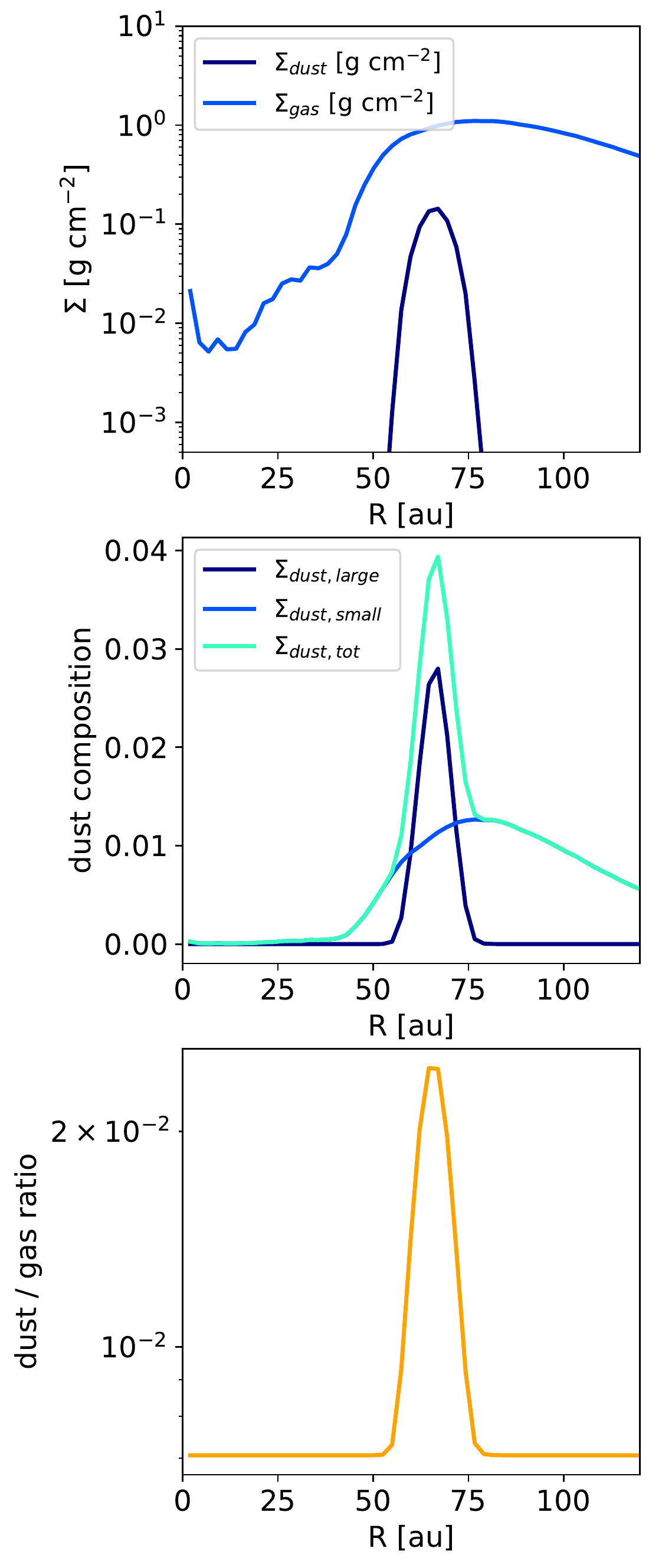}
 \caption{Top: surface density profiles of gas (light blue) and dust (blue) plotted as a function of the radius for the t = 500 T$_{\rm orb}$ ($\sim$ 0.18 Myr) snapshot of Sim~3. Middle: Profiles of the two synthetic dust populations (large and small grains) and of the total dust distribution. Bottom: dust-to-gas ratio any given point.}
 \label{fig:fig_dust_to_gas}
\end{figure}

\begin{figure*}
 \includegraphics[width=\textwidth]{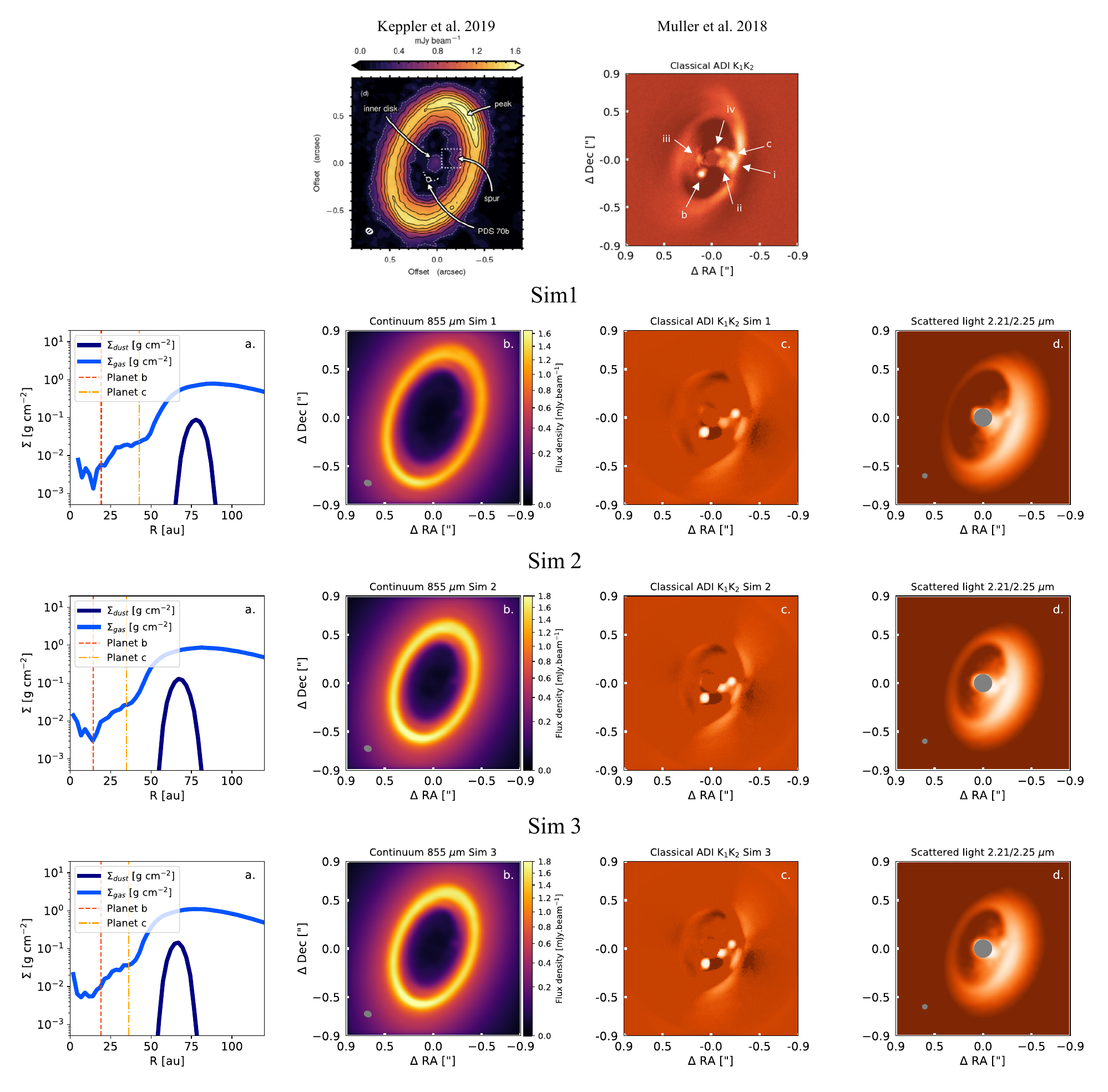}
 \caption{Comparison between our three simulations (bottom three rows) and observations from \citet{Keppler2019} and \citet{Muller2018} (top row). Panels a show the surface density profile of gas (light blue) and dust (blue) plotted as a function of radius for the t = 500 T$_{\rm orb}$ snapshot for the three simulations (1 to 3 from top to bottom), in units g cm$^{-2}$ and au respectively. Instantaneous planets positions are also plotted as red and orange vertical lines. Panels b show the corresponding simulated continuum image at 855 $\mu$m, with fluxes in Jy beam$^{-1}$. The image has been smoothed with a 74 $\times$ 53 mas Gaussian beam to match the beam size of \citep{Keppler2019}. Panels c and d show the average between the 2.11/2.15 $\mu$m scattered light images with and without median-ADI processing, respectively. In panels c the flux of the planets were injected in the ADI cube before processing, while panels d do not include emission/heating from the planets. The flux density is in Jy/arcsec$^{-2}$ and the image has been convolved with a psf of 0.05 mas.}
 \label{fig:fig_1}
\end{figure*}

\section{Results}\label{sec:Results}
Figure~\ref{fig:fig_surf_gas} shows the gas and dust (1 mm) surface density distribution for the t = 500 T$_{\rm orb}$ ($\sim$ 0.18 Myr) snapshot of Sim~3. The presence of the two planets opens a large cavity in the gas, while the dust is trapped in a ring due to the combined effect of dust trapping in the pressure maximum \citep{Pinilla2012} and radial drift. This is consistent with the results of \citet{Bae2019}, where only smaller ($ < \mu m$) dust grains can filtrate inside the cavity. As visible in Fig.\ref{fig:fig_surf_gas}, an accretion stream connects Planet c with the gas rim of the disc, highlighting that the planet is actively interacting with the disc.

In our simulations, a circumplanetary disc is created around all the planets and quickly (in about 50-100 T$_{\rm orb}$) accreted. In Figure~\ref{fig:fig_surf_gas}, shown after 500 T$_{\rm orb}$, only gas around the planets can be spotted.

Figure~\ref{fig:fig_dust_to_gas} shows how the synthetic dust grain population is obtained from the simulation during the post-processing with MCFOST. In Fig.~\ref{fig:fig_dust_to_gas} (top) the surface density profiles of gas (light blue) and dust (blue) are plotted as a function of the radius for the above snapshot, in Fig.~\ref{fig:fig_dust_to_gas} (middle) we show the density profiles of the two synthetic dust populations (large and small grains) and of the total dust distribution. Finally,  Fig.~\ref{fig:fig_dust_to_gas} (bottom) displays the local value of the dust-to-gas ratio any given radial point.

\subsection{Synthetic Images}\label{sec:SynthImgs}
Figure~\ref{fig:fig_1} shows the surface density profile of gas (light blue) and dust (blue) plotted as a function of the radius for the t = 500 T$_{\rm orb}$ snapshots (Panels~a). For each simulation, we also show the simulated dust continuum image at 855~$\mu$m convolved with a 74 $\times$ 57 mas Gaussian beam (Panels~b), and the average between the 2.11~$\mu$m and 2.25~$\mu$m scattered light image, either obtained after the steps (i) to (ix) detailed in Section~\ref{IRpostprocessing} (Panels~c), or only convolved with a PSF of 0.05 mas (Panels~d). The latter aim to show the authentic morphology of scattered light signals before ADI processing. Table~\ref{table:tab2} lists the masses and positions of the planets as well as the disc mass at the time of the snapshots. 

All our models reproduce the observed morphology of the dust and gas rings, as well as both the small features that are found in the images (the dust skewness seen from \citealt{Isella2019} and \citealt{Keppler2019} and most of the features seen in the cavity in the 2.11/2.15$\mu$m SPHERE images from \citealt{Muller2018}).

% Best dump planets table
\begin{table}
    \centering
    \caption{Masses and positions of the two planets, gas and dust disc masses for our set of simulations at t=500 T$_{\rm orb}$. Planets' masses are in Jupiter mass (M$_{\rm J}$), distances are in au and the disc masses are in Solar masses.}\label{table:tab2}
    \begin{tabular}{*4c}
        \toprule
        & \multicolumn{3}{c}{Simulations} \\
        \cmidrule(lr){2-4}
               & Sim 1  & Sim 2  & Sim 3 \\    
%        Name   &        &         &         \\
        \midrule
        $M_{\rm b}$ ($M_{\rm J}$)     & 4.4 & 5. & 5.3   \\ 
        $R_{\rm b}$ (au)        & 19.4  & 14.2  & 19.1  \\
        $M_{\rm c}$ (M$_{\rm J}$)     & 4.1 & 4.3 & 3.9 \\
        $R_{\rm c}$ (au)        & 42.6  & 34.9  & 36.1  \\
        $M_{\rm gas}$ (M$_\odot$) &  2.5 $\times$ 10$^{-3}$ & 3.5 $\times$ 10$^{-3}$ & 3.3 $\times$ 10$^{-3}$\\
        $M_{\rm dust}$ (M$_\odot$) &  5.4 $\times$ 10$^{-5}$  & 6.9 $\times$ 10$^{-5}$ & 7.0 $\times$ 10$^{-5}$ \\
        \bottomrule
    \end{tabular}
\end{table}

Focusing on the surface density profiles (Panels a in Fig.~\ref{fig:fig_1}), the position of the inner Planet~b (plotted as an orange vertical lines) does not appear to impact the shape of the surface density profile: all the simulations show a large and deep cavity in the inner --- $\sim$~50~au --- part of the disc (the surface density in the inner part is at least two orders of magnitude lower than the outer part of the disc), regardless of the final positions of Planet~b, that migrates very little in all cases. On the contrary, the final position of the outer planet (Planet~c, shown as a yellow vertical line) affects both the width of the gas cavity (larger for larger values of the initial Planet~c distance, as in Sim~1) and the position of the dust ring (that appears at $\sim$ 65--90 au for Sim~1 and at $\sim$ 50--75 au in the other cases). 

For this reason, the 855 $\mu$m image of Sim~1 has a fainter and larger ring compared to the images of Sim~2 and 3 (see panels b in Figure~\ref{fig:fig_1}). However, a spur is always present, in the same position, in all cases. This is in agreement with what observed with ALMA \citep{Isella2019}. This feature corresponds exactly to the location of Planet~c, and is due to the interaction between the planet and the outer part of the disc. Indeed, the planet is actively accreting material from the dust and the gas ring that may form a circumplanetary dust ring. This feature has already been reproduced and analysed in \citep{Bae2019}.

\begin{figure}
\centering
 \includegraphics[width=\columnwidth]{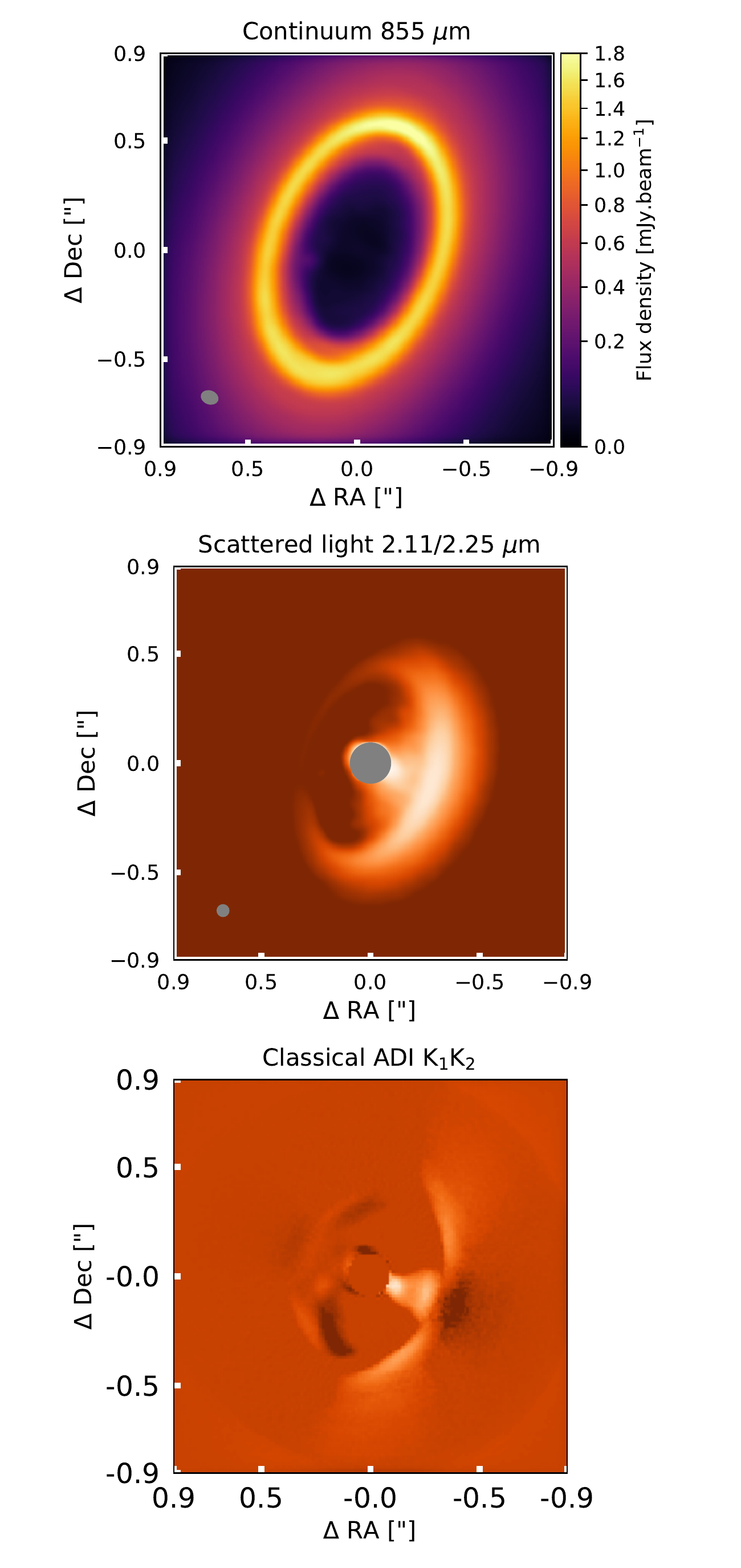}
 \caption{Simulated continuum image at 855 $\mu$m (top) and the average between 2.11/2.15 $\mu$m scattered light images (bottom) for the  t = 500 T$_{\rm orb}$ snapshot of sim 3 when planet c is located 180deg forward in true anomaly, compared to the observed one. While the forward-scattered intra-cavity dust signal remains in the scattered light image, the spiral accretion stream, including the bright blob surrounding c, disappears.}
 \label{fig:fig_2}
\end{figure}

Panels~c and d of Figure~\ref{fig:fig_1} show the synthetic 2.11/2.15 $\mu$m images of our simulations with and without median-ADI post-processing, respectively. Contrary to panels~d, panels~c were obtained with the flux of planets b and c injected in the synthetic ADI cube at the location of the sink particles. This is for a fairer comparison to the observations, given that the bright flux of the planets also affects the median-ADI algorithm. Sim~1 can be seen to produce a larger cavity and a broader ring compared to the other two cases, as already discussed. However, we stress that we naturally explain most of the observed bright features in the scattered light image without any specific assumption. 
In order to interpret the origin of these features, labelled \textit{i} to \textit{iv} in Figure~\ref{fig:fig_1}, we also processed
a snapshot of the simulation corresponding to planet c located at the opposite point of its orbit (+180 degrees true anomaly) compared to the observations. This is shown in Figure~\ref{fig:fig_2} for Sim 3.

The dust spur (seen in the 855 $\mu$m image, Panels b) is now present on the other side of the disc, confirming that this is not a numerical feature but is due to the presence of dust surrounding Planet c. However, some of the intra-cavity emission features in the 2.11/2.15 $\mu$m image (panels c) are still present at roughly the same location. Features \textit{i} to \textit{iv} are interpreted as follow:
\begin{enumerate}
    \item Feature \textit{i} is an artefact of median-ADI processing of the bright forward-scattered edge of the cavity. Similar artefacts were obtained in the forward modelling tests shown in \citet{Milli2012}. 
    \item Feature \textit{ii} likely consists of two contributions: scattered light signal from small intra-cavity dust (as suggested from the blob at the same location in the bottom panel of Figure~\ref{fig:fig_2}) and from the spiral accretion stream extending inward from c. The latter is confirmed by the residual blob seen at the location of feature \textit{ii} in the image obtained when subtracting the scattered light snapshot where planet c is 180deg away from its observed location from the snapshot where it is at its observed location (Appendix \ref{app:SpiralStream}).
    \item Feature \textit{iii} could trace the back-scattered emission from small dust grains in the cavity, as a faint signal is also seen on that side of the cavity in our predicted images.
    \item Feature \textit{iv} is not reproduced in our processed RT images. It is unclear whether this point-like feature discussed in \citet{Mesa2019} could trace the edge of a denser inner disc -- not  present in our simulations -- or whether it is a residual speckle.
\end{enumerate}
A more detailed discussion of post-processing artefacts including a comparison between polarimetric differential imaging (PDI), median-ADI and principal component analysis (PCA) ADI observations on the one hand, and our predicted processed images in the other is provided in Appendix \ref{app:DIeffects}. 

In all our images the dusty feature observed around the inner planet (Planet~b) in \citealt{Isella2019} is absent. This does not mean that a dusty circumplanetary disc should not be present around this planet. Indeed, in our simulations dust material surrounding this planet forms a disc that is quickly accreted on the sink due to numerical effects. For a more complete discussion on this feature see \citealt{Bae2019}.

\begin{figure*}
\centering
 \includegraphics[width=0.9\textwidth]{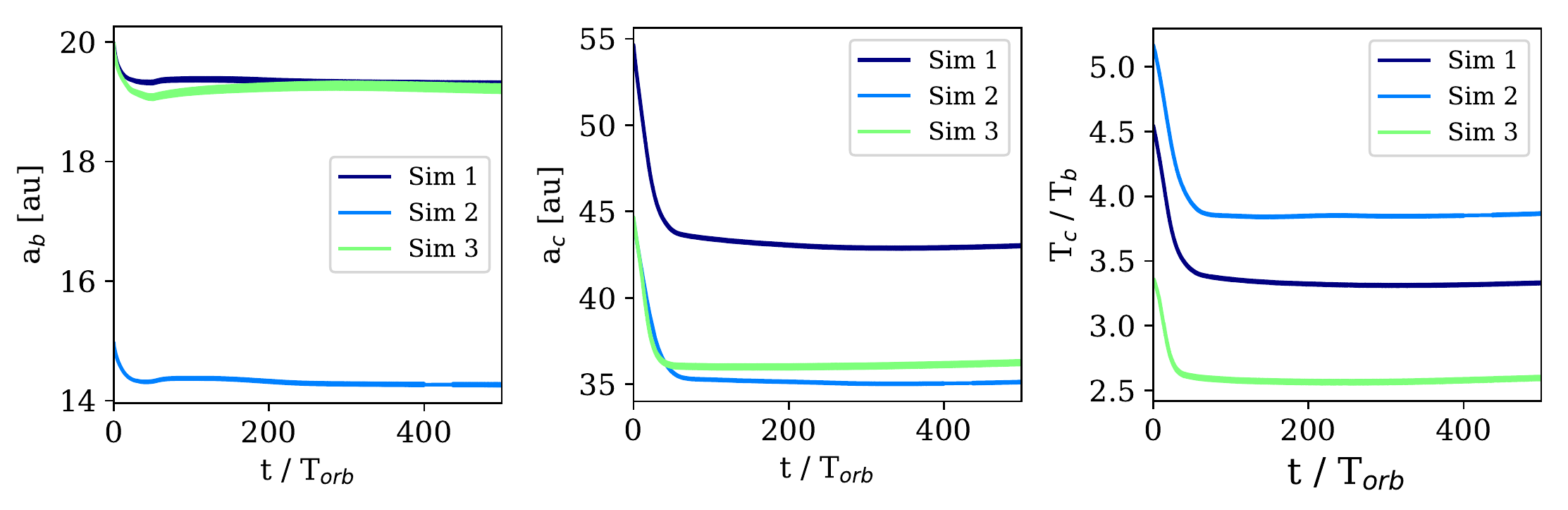}
 \caption{Planet positions for the simulations: Semi-major axis of Planet~b (left), Planet~c (middle) and the ratio between the two orbital periods (right) plotted as a function of the normalized time for the three simulations (see legend). The period ratio indicates that planets reach resonances like 5:2 and 7:2.}
 \label{fig:fig_7}
\end{figure*}

\begin{figure*}
\centering
 \includegraphics[width=0.9\textwidth]{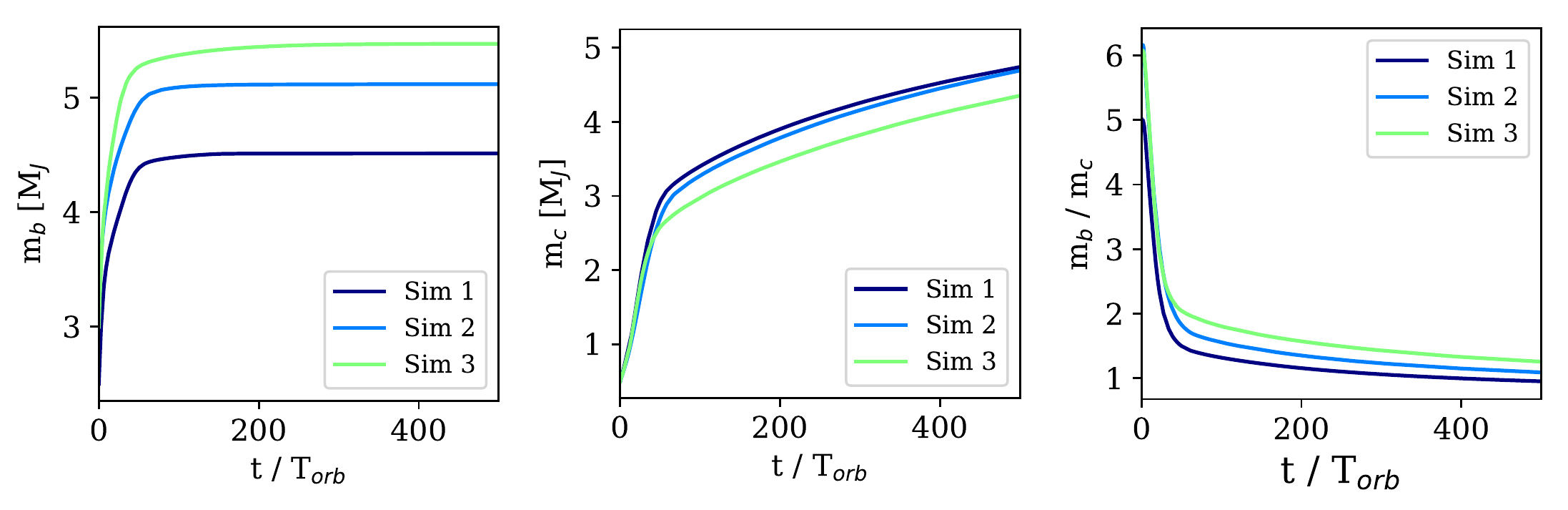}
 \caption{Masses of Planet b (left) and Planet c (middle) as a function of the normalized time, and their relative ratio (right) for the three simulations.}
 \label{fig:fig_6}
\end{figure*}

\subsection{Overall dynamics}

Figures~\ref{fig:fig_7} and \ref{fig:fig_6} illustrate the dynamics of the planets for the three simulations. In particular, Figure~\ref{fig:fig_7} shows the position of Planet~b (left), Planet~c (middle) and the ratio between the two orbital periods (right), to check whether or not they settle into a mean motion resonance, as a function of the normalized time. 
In the first $\sim$ 50 orbits the systems relaxes out of the initial conditions: Planet~c quickly migrates $\sim$~10~au inward (middle panel), while the inner planet (Planet~b) remains closer to its initial position, due to its larger initial mass (left panel). The migration of the outer planet ends when the orbits of the two objects lock in a period resonance
(right panel) that is different in the various simulations. This value appears to be dependent on the initial position of the outer planet (Planet~c): a larger initial distance between the star and Planet~c (Sim~1) leads to an equilibrium configuration when Planet~c more distant compared with the other cases (Sim 2 and 3). No significant dependence  on the initial position of the inner planet (Planet~b) is found (see the behaviour of Sim 2 and Sim 3).
The value of the ratio between the orbital periods depends also on the position of the inner planet, Planet~b. In order to match the 2:1 expected resonance for this system as well as the ALMA observed 855 continuum flux, Planet~b should be located at $\sim$17--22~au and Planet~c at $\sim$34--39~au. This situation is most closely matched by Sim~3. 

\begin{figure*}
\centering
 \includegraphics[width=\textwidth]{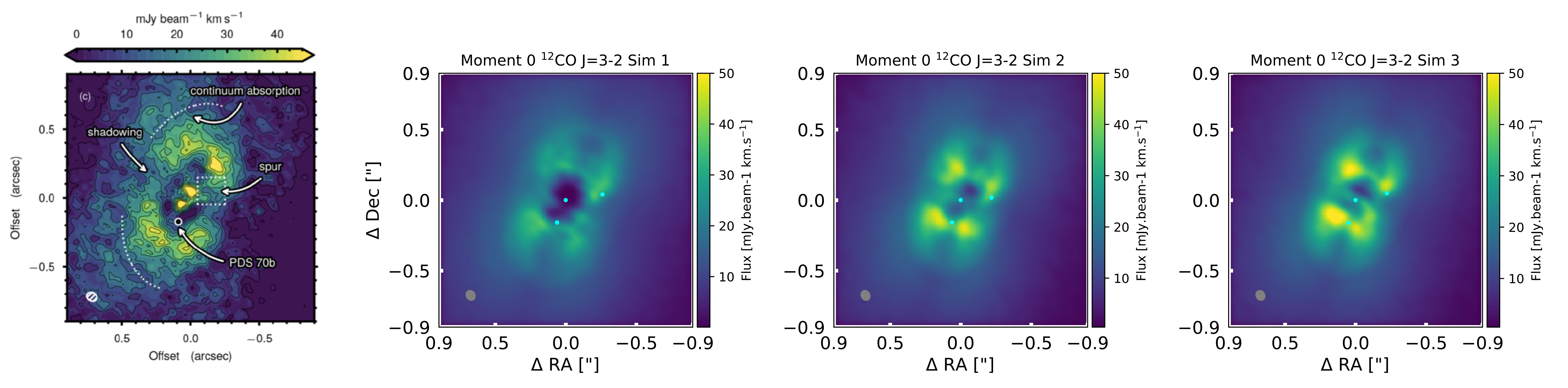}
 \caption{CO J=3-2 integrated intensity (moment 0, continuum subtracted) for all the simulations for the t = 500 T$_{\rm orb}$ snapshot and, as a reference, the observed image of \citet{Keppler2019} (left). The position of the star and the planets is shown as cyan dots. The image is convolved with a Gaussian 76 $\times$ 61 mas beam with PA of 63 degrees as in \citet{Keppler2019} observations.}
 \label{fig:fig_mom}
\end{figure*}

Figure~\ref{fig:fig_6} shows the masses of Planet~b (left) and Planet~c (middle) as a function of the normalized time, and their relative ratio (right). After the
first $\sim $ 100 T$_{\rm orb}$ Planet b has already accreted almost all of the available mass, reaching the value of 4.5 - 5 M$_{\rm J}$, depending on the initial conditions, while the outer planet continues to accrete mass from the external (R $>$ 50 au) reservoir of gas, reaching $\sim$ 5 M$_{\rm J}$ mass at the end of the simulation, with a significant mass accretion rate. This suggests that at the end of the evolutionary path of the disc, the mass of the outer planet could be even larger.
In all our models the orbital eccentricity of the planets remains relatively small, less than about 0.1 for Planet c (with an exact value dependent on the planet mass) and $\sim$ 0.15 for Planet b, values compatible with observations and numerical results from previous simulations \citep{Bae2019}. 

No dust ring between the two giants is observed in the mm continuum observations. In our simulations an inter-planetary ring forms in the first $\sim$~100 orbits but it is quickly accreted about $\sim$~100--150 orbits later, showing that in this case, even if a non-negligible amount of dust is present at the beginning of the simulations, the dust is not trapped between the orbits of the planets, but it is accreted onto the sinks.
\begin{figure*}
\centering
 \includegraphics[width=\textwidth]{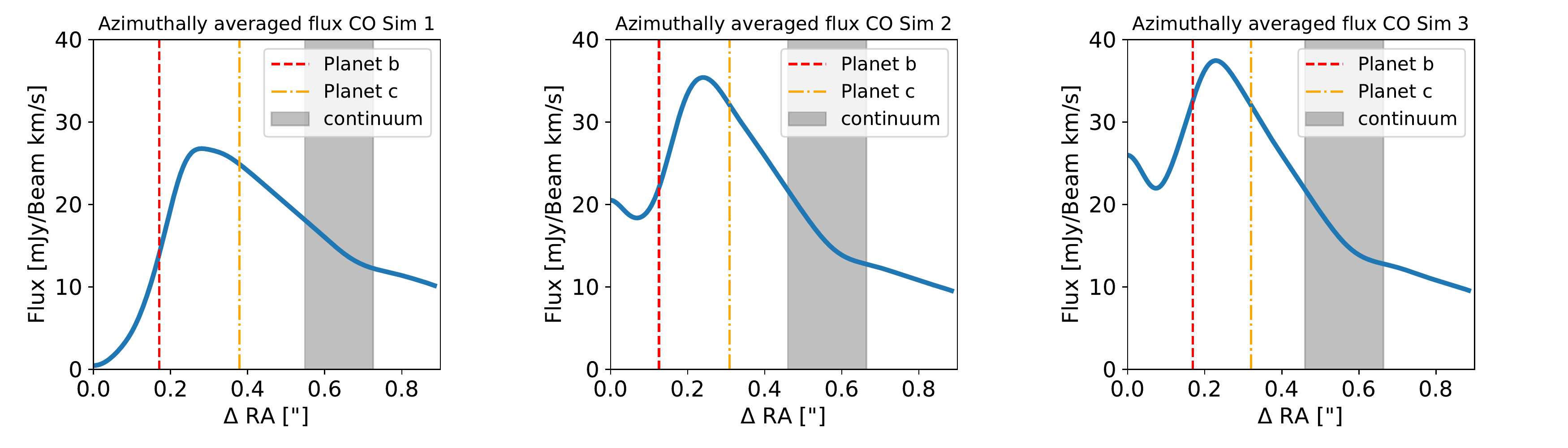}
 \caption{Azimuthally averaged $^{12}$CO moment zero maps (continuum subtracted) for for all the simulations (left: Sim 1, middle: Sim 2, right: Sim 3) for the  t = 500 T$_{\rm orb}$ snapshot. The position of Planet b and c is shown as the red and orange vertical line respectively, while the grey region shows the extent of the continuum ring.}
 \label{fig:fig_azim}
\end{figure*}

\subsection{CO maps and mass accretion rate}

In order to test whether our models are compatible with CO observations of \citet{Keppler2019} we also produced synthetic CO moment zero maps. Figure~\ref{fig:fig_mom} shows the synthetic emission in $^{12}$CO J = 3 - 2 of our three models (last three panels) compared to the observations (left panel). The two planets open a $\sim 40$ au wide cavity with a decrease in surface density of at least two orders of magnitude in the gas (see Fig.\ref{fig:fig_1}), with an emptier inner region (R < 25 au). This results in the CO moment zero map in an inner gap with no emission (all the gas is photo-dissociated) and the presence of an asymmetric ring at $\sim$ 0.03 arcsec, asymmetry probably due to the presence of high optical depth zones. The flux value is compatible with the observed values. 
Figure~\ref{fig:fig_azim} presents the azimuthally averaged CO deprojected flux density of our models. In each panel, the grey area shows the position of the continuum ring, while the vertical red and orange line corresponds to the position of PDS~70b and c. As in \citet{Keppler2019}, the peak of the CO emission is located approximately close to the position of Planet~b, about 0.3 arcsec, but we also show the position of Planet~c, showing that CO material is present in the neighbourhood of the planets' orbits.  We also find a change in the slope of the profiles located at 0.6~arcsec. Comparing our images with \citet{Keppler2019} observations, we observe that the the observed CO images peaks at about 0.4 arcsec while the models show the peak at about 0.3~arcsec. This difference is probably due to a higher gas mass inside the cavities in our models with respect to the source, accentuated by the absence of an inner ring that can shield the material from the radiation of the star. Higher resolution models, or simply evolving the simulations for longer may help to resolve this discrepancy.

\begin{figure*}
\centering
 \includegraphics[width=0.9\textwidth]{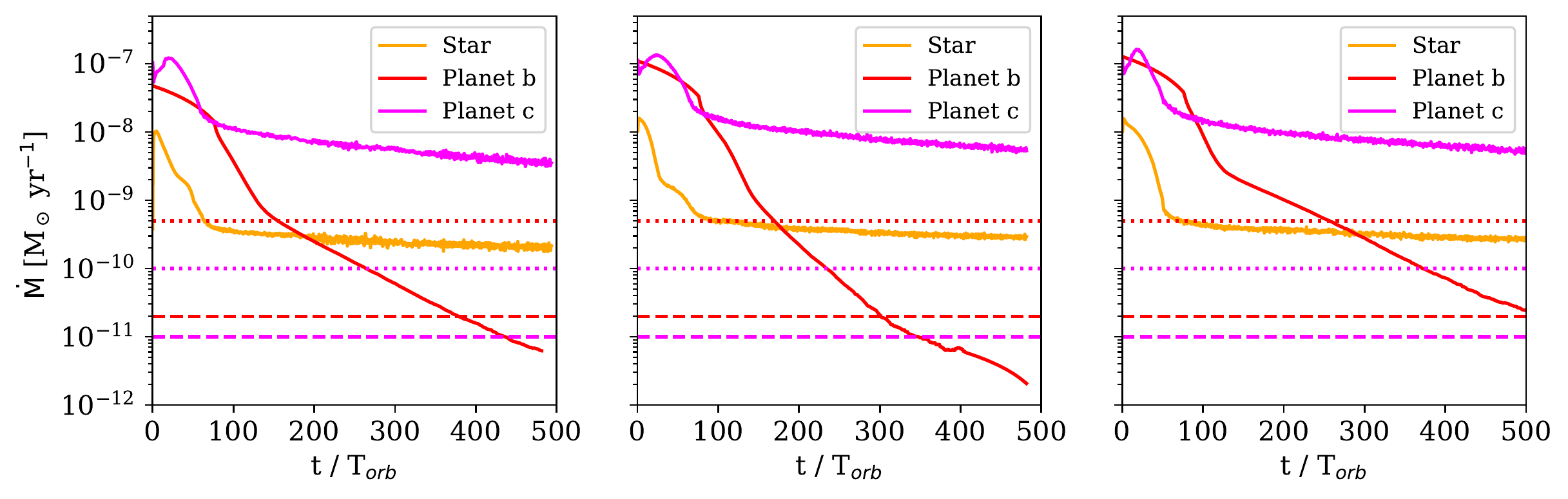}
 \caption{Accretion rates (in M$_\odot$ yr$^{-1}$) onto the sinks for all the simulations (left: Sim 1, middle: Sim 2, right: Sim 3) as a function of the normalized time. Orange lines are the star accretion rate and red and pink lines are Planet b and Planet c accretion rates respectively. Horizontal red and pink dashed lines show the inferred values from Planet b and c respectively from \citet{Haffert2019}, while the horizontal red and pink dotted lines show the inferred values from Planet b and c from \citet{Hashimoto2020}.}
 \label{fig:fig_3}
\end{figure*}

Figure~\ref{fig:fig_3} shows the mass accretion rates (in M$_\odot$ yr$^{-1}$) onto the planets in our simulations (left: Sim 1, middle: Sim 2, right: Sim 3) as a function of normalized time. 
We recall that by definition in SPH all the material accreted by the sink is merely added to the planet mass. In reality, part of the ``accreted mass'' may end up in an unresolved circumplanetary disc that buffers the material onto the planet, so that our computed accretion rates may be considered as upper limits.

All three simulations show the same behaviour: the accretion rate on the star is $\sim 5 \times 10^{-10}$ M$_\odot$ yr$^{-1}$, an order of magnitude higher than the observed value. The fact that our sink radius for the star (2 au) is far larger than the actual radius for the star can be the reason for this overestimate. However, after relaxing out the initial conditions, this value slowly decreases with time, suggesting that gas is continuously being accreted on the star. Our simulations also show that the outer Planet~c has a higher accretion rate ($ \sim 10^{-8}$ M$_\odot$ yr$^{-1}$) with respect to the inner Planet~b, ($ \leq 10^{-10}$ M$_\odot$ yr$^{-1}$). This occurs naturally in our simulations because the outer planet is directly in contact with the external mass reservoir, in contrast to the inner planet, that sits within the cavity.

Moreover, while the accretion rate of Planet c is (almost) constant with time, because it is actively accreting material (as the morphological features --- seen in dust and gas --- emphasize), the accretion rate of Planet b rapidly drops as the cavity is depleted, "starving" the inner planet.  This occurs because the outer giant planet prevents the replenishment of the inner disc.

The accretion rate we obtain for Planet~b is consistent with the observed value of $\sim 2 \times 10^{-11} M_{\sun}$ yr$^{-1}$. By contrast, we appear to strongly overestimate the accretion rate onto Planet~c (which is $ \times 10^{-11} M_{\sun}$ yr$^{-1}$). As mentioned above, we stress that the ``accretion rate'' we measure is actually a measure of how much mass enters the Hill's sphere of the planet. Most likely, a circumplanetary disc would form (which would be unresolved in our simulations) which would act as a buffer for the accretion flow onto the planet. We also neglect any radiative feedback from the planets. Without a proper modelling of such disc, we cannot draw any firm conclusions regarding the planetary accretion rates. However, regardless of the absolute values of the accretion rates, their relative ordering, with the outer planet accreting at a higher rate than the inner planet, should be preserved and is a natural consequence of the inner disc depletion. 

\section{Discussion}\label{sec:Discussion}

Our simulations reproduce the main features found in ALMA 855 $\mu$m and VLT/SPHERE 2.11/2.15 $\mu$m observations (see Fig. \ref{fig:fig_1}). In particular, the complex morphology observed in scattered light at the inner edge of the ring is explained by a combination of post-processing biases and the high inclination of the disc, with a likely minor contribution from a spiral accretion stream (Section~\ref{sec:SynthImgs} and Appendix~\ref{app:SpiralStream}).

We find that the best scenario (Sim~3) gives final values for the planet masses of $\sim 5-10$ M$_{\rm J}$, with Planet b located at $\sim$ 18-20 au and Planet c located at $\sim 35-40$ au, in a 5:2 mean motion resonance, in agreement with observational results \citep{Keppler2019} and previous models \citep{Bae2019}.

The initial mass of the inner planet (Planet b, 2.5-3 M$_{\rm J}$) is lighter than that of the outer planet (Planet c, 0.5 M$_{\rm J}$). This could either represent a situation where the two planets are born at a comparable time with very different masses or the effectively equivalent case where the inner planet forms earlier, migrates and reaches a higher mass at the time where the second planet forms in the outer disc. We show that giant planets can reach an orbital resonance while they still are embedded in the disc and still accreting gas, supporting \citet{Bae2019}'s findings for PDS 70. The actual value of the resonance appears to depend on the disc and planet masses, but further studies are necessary to better characterize this phenomenon. This behaviour has also been proposed for Jupiter and Saturn in the Solar System (see e.g \citealt{Masset2001}), with the caveat that PDS 70 and our Solar System are very different due to the huge differences in the giant planet masses and positions. 

The presence of two giants planets on stable orbits sculpts the host disc, opening a wide cavity in the dust and gas density and blocking the gas reservoir and the mm dust ring in the outer part of the disc. Filtration mechanisms could be at play, as found in \citet{Bae2019}: large grains are trapped beyond their orbits, while smaller grains can filtrate, changing the amount of solids available in the inner and outer part of the disc, impacting on the formation of terrestrial planets and possibly on the composition of the cores of the giants. Indeed, this can eventually lead to the formation of two different and spatially separated mixtures of dust and gas, that may have influences on the future planetary formation and chemical composition. Focusing on the gas, the flux values of the synthetic CO moment zero maps are compatible with the observations of \citet{Keppler2019}, however we fail to reproduce the compact inner ring found in the observations, probably because of our use of a 2~au sink radius for the star. Considering the azimuthally averaged flux profiles, the ring we find has the correct flux value but is located closer to the star (0.3 arcsec in our models and 0.4 arcsec in the observations), pointing out that probably there is too much gas in the cavity. However, finite resolution effects could be at play, due to the fact that the planets are actively emptying the cavity, decreasing also the SPH resolution.

When two giant planets in resonance shape their host disc, the relative distance between their positions appears to be particularly relevant to determine the dust morphology. Indeed, in the case of PDS 70, where the orbital ratio is close to 2:1 no dust ring is observed between the orbits of the planets. 
By contrast, in the case of HD 169142, another source where at least two candidate giants protoplanets could be present \citep{Fedele2017,Perez2019}, a prominent dust ring is observed between the likely position of the two giants, that are expected to be in a larger orbital ratio. Numerical results \citep{Toci2020} confirm the presence of a long lived dusty ring in this case. Apparently, whether or not a dust ring between the two planets is present depends on the order of the mean-motion resonance, with closer orbits not allowing a long-lived inter-planetary dust ring.

Another result from our simulations is that the relative position between the planets and the dust or gas material in the disc plays a key role in inferring their presence. While ultimately only a direct measurement of a planet can confirm its presence, detection of accretion streams or sub-mm circumplanetary discs can help pinpoint the candidate planet location.

It is unclear whether the observed mm-continuum emission near c (refereed to as ``spur'') corresponds to circumplanetary disc emission, or whether it traces the wake from c passing through that location. The latter interpretation would be consistent with all the non-detections of circumplanetary discs \citep[e.g.][]{Ricci2017}, the sub-mm image of PDS 70 shown in \citet{Isella2019} after subtraction of the symmetric component (showing a stream-like residual), and the sub-mm spiral-like feature seen within an annular gap in the HD~163296 disc \citep{Isella2018}. 
The shift between the IR and sub-mm signals near b could possibly be explained in a similar way. 

One of our aims was to test if the results of our simulations are sensitive to the initial conditions for the planets. In the small sample shown, the initial positions of the planets, as well as their initial masses, appear not to have a large impact on the global shape of the disc. However, a more detailed study of the parameter space is needed to understand the robustness of the model

Our simulations did not allow us to firmly constrain the planetary accretion rates.
Numerical effects, due to a high numerical viscosity given by low resolution of the circumplanetary disc around the sink particles, would lead to a shorter life-time for the circumplanetary disc, drastically increasing the mass accretion rate. Particularly sensitive to these effect is the case when the sink is continuously interacting with a reservoir of material, as in the case of Planet c.

However, we do expect that the relative ordering of the two planetary accretion rates found in our simulations should be preserved at higher resolution, with the outer planet accreting at a higher rate. Current estimates of the planetary accretion rates in PDS~70 \citep{Haffert2019,Hashimoto2020} appear to go in the opposite direction, although these are affected by large uncertainties.

\section{Conclusions}\label{sec:conclusions}
In this work we performed 3D SPH simulations of the PDS~70 protoplanetary disc. Because two giant planets have been directly observed in this source, we initialised our simulations with two planets embedded in a disc of dust and gas, with smaller initial mass and larger orbital ratio than observed. In contrast to previous modelling attempts, we allowed our planets to grow in mass and to migrate within the disc.
Our simulation setup mimics a scenario in which the inner and the outer planets form and grow at different locations in the disc and are captured in a mean-motion resonance while migrating and accreting mass.
We found that the two planets quickly lock into resonance with a ratio that is compatible with observational results as well as with previous 2D numerical simulations \citep{Bae2019}. We also found excellent agreement between our mm and scattered-light images and the literature, reproducing the main observed features caused either by the presence of the outer planet or by geometric biases related to high-contrast IR observations and post-processing.

Planet-disc and planet-planet interactions are at play in PDS~70, carving the disc in the dust and gas reservoir. Planet migration and the resonant locking of the orbits are key processes. Millimetre-emitting dust grains trapped in a gas  pressure bump at $\sim$75 au cause the observed ring seen in continuum emission, supporting the results from \citealt{Bae2019}.
\section*{Data availability}
The data underlying this article will be shared on reasonable request to the corresponding author.

\section*{Acknowledgements}
The authors want to acknowledge the referee and the Editor for the useful suggestions.
This work and CT are supported by the PRIN-INAF 2016 The Cradle of Life - GENESIS-SKA (General Conditions in Early Planetary Systems for the rise of life with SKA). 
CT, DF and LT acknowledge financial support provided by the Italian Ministry of Education, Universities and Research. CT and LT through the grant Progetti Premiali 2012 – iALMA (CUP C52I13000140001), DF from project SIR (RBSI14ZRHR); LT aknowledges the Deutsche Forschungsgemeinschaft (DFG, German Research Foundation) - Ref no. FOR 2634/1ER685/11-1 and the DFG cluster of excellence ORIGINS (www.origins-cluster.de).
GL and LT received funding from the EU Horizon 2020 research and
innovation programme, Marie Sklodowska-Curie
grant agreement 823823 (Dustbusters RISE project). 
VC, CP and DJP acknowledge funding from the Australian
Research Council via FT170100040 and
DP180104235 and from ANR of France (ANR-16-CE31-0013).
This paper makes use of the following ALMA data: ADS/JAO.ALMA 2013.1.00592.S. ALMA is a partnership of ESO (representing its member states), NSF (USA) and NINS (Japan), together with NRC (Canada), MOST and ASIAA (Taiwan), and KASI (Republic of Korea), in cooperation with the Republic of Chile. The Joint ALMA Observatory is operated by ESO, AUI/NRAO and NAOJ.
Computational resources provided by INDACO Platform, a project of High Performance Computing at the Università degli Studi di Milano http://www.unimi.it; 
CT thanks Eleonora Bianchi, Benedetta Veronesi, Simone Ceppi and Giovanni Poggiali for the useful discussions and Prof. S. Nayakshin for improving the quality of this work.
%%%%%%%%%%%%%%%%%%%%%%%%%%%%%%%%%%%%%%%%%%%%%%%%%%

%%%%%%%%%%%%%%%%%%%% REFERENCES %%%%%%%%%%%%%%%%%%

% The best way to enter references is to use BibTeX:

\bibliographystyle{mnras}
\bibliography{paper_mnras_template} % if your bibtex file is called biblio.bib

% Alternatively you could enter them by hand, like this:
% This method is tedious and prone to error if you have lots of references
%\begin{thebibliography}{99}
%\bibitem[\protect\citeauthoryear{Author}{2012}]{Author2012}
%Author A.~N., 2013, Journal of Improbable Astronomy, 1, 1
%\bibitem[\protect\citeauthoryear{Others}{2013}]{Others2013}
%Others S., 2012, Journal of Interesting Stuff, 17, 198
%\end{thebibliography}

%%%%%%%%%%%%%%%%%%%%%%%%%%%%%%%%%%%%%%%%%%%%%%%%%%

%%%%%%%%%%%%%%%%% APPENDICES %%%%%%%%%%%%%%%%%%%%%

\appendix

\section{Spiral accretion stream in infrared images}\label{app:SpiralStream}

\begin{figure}
\centering
 \includegraphics[width=\columnwidth]{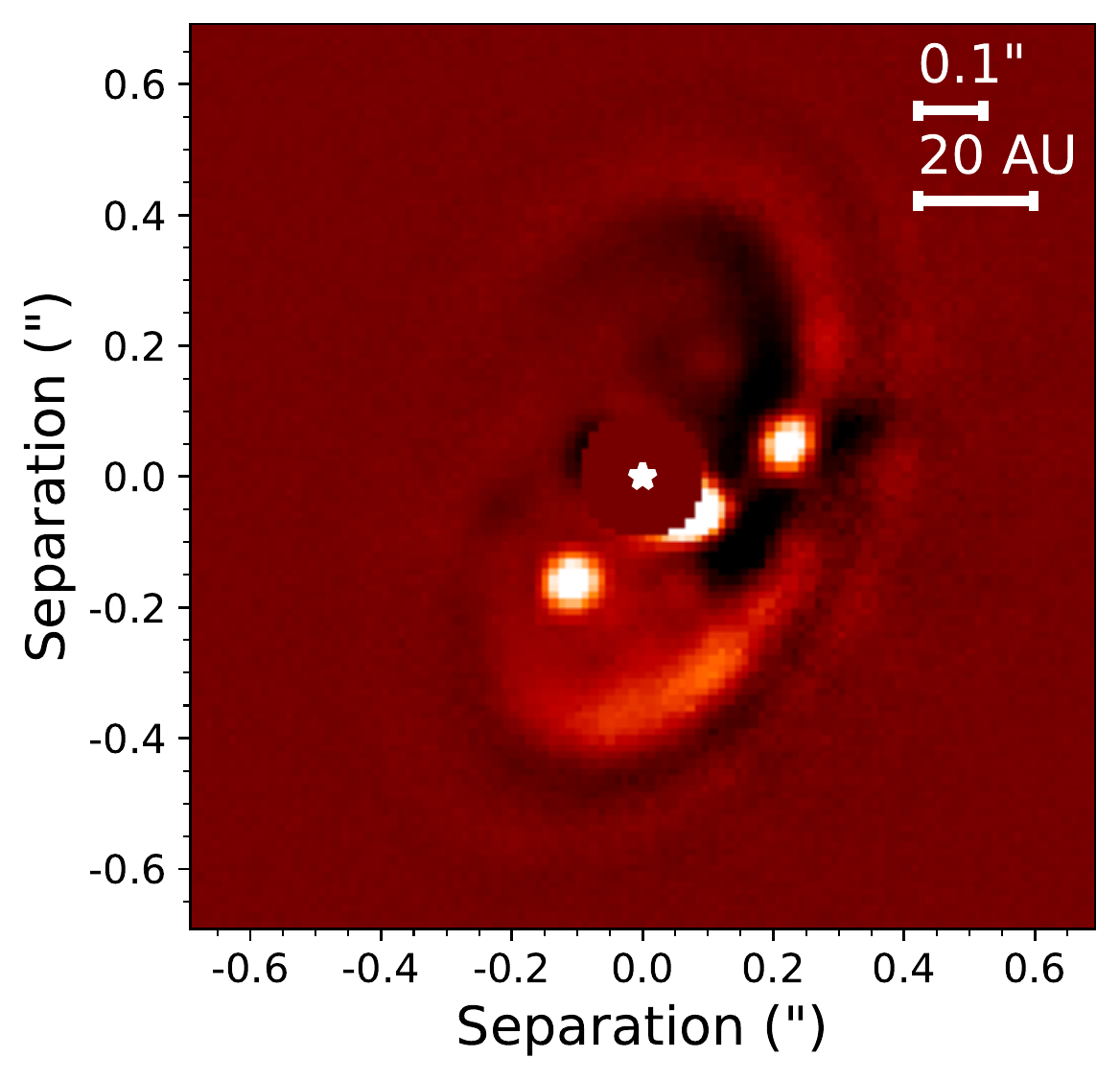}
 \caption{Differential image obtained by subtracting the predicted scattered light image at 2.11/2.25 $\mu$m when the location of planet c is 180 degrees from the observed one to the predicted image where the planets are at their observed position. This highlights the faint spiral accretion stream induced by planet c.}
 \label{fig:Diff180deg}
\end{figure}

Given both the biases induced by angular differential imaging (ADI; Figure~\ref{fig:fig_1}, third column) and the presence of small dust grains leading to bright intra-cavity signals regardless of planet c's position (Figure~\ref{fig:fig_2}), it is unclear which scattered light signals result from the interaction of planet c with the disc. In order to highlight this specific signature we show in Figure~\ref{fig:Diff180deg} the image obtained when subtracting the scattered light image with planet c located 180 deg away from its observed position (and no planet flux injected) to the scattered light image obtained when the planets are injected at their observed position angle and contrast.
These images are obtained considering convolution with the observed point spread function and the effect of a coronagraph, but without geometric bias induced by ADI.
The faint spiral accretion stream associated to c can be seen both outward and inward from its location. In particular, the blob that is found inward suggests that part of the signal from Feature \textit{ii} (Figure~\ref{fig:fig_1}) could trace the inner spiral arm of PDS 70 c.
It is worth noting that the spiral arm seen along the edge of the cavity at the bottom part of the image is not a secondary arm but results from the subtraction of the primary spiral associated to planet c when it is placed at 180deg difference in true anomaly compared to observations.

Spiral arms have been observed in scattered light observations in a number of protoplanetary discs \citep[e.g.][]{Muto2012,Benisty2015,Uyama2020}, and have been interpreted as possible planet signposts \citep[e.g.][]{Dong2015,Zhu2015}. However they have not been reported in this system, despite the presence of confirmed embedded protoplanets. 
The test presented in this appendix suggests that the lack of reported spirals is most likely due to the unfavourable viewing geometry of the disc, lack of angular resolution in the deprojected radial direction and faint additional flux compared to the bright forward-scattered edge of the cavity.
Our conclusion meets the one of \citet{Zhu2015} who showed that a high inclination can significantly hinder the visibility of spiral arms, in particular in polarised light due to the phase function of the degree of polarisation.

\section{Differential imaging effects}\label{app:DIeffects}

\begin{figure*}
\centering
 \includegraphics[width=\textwidth]{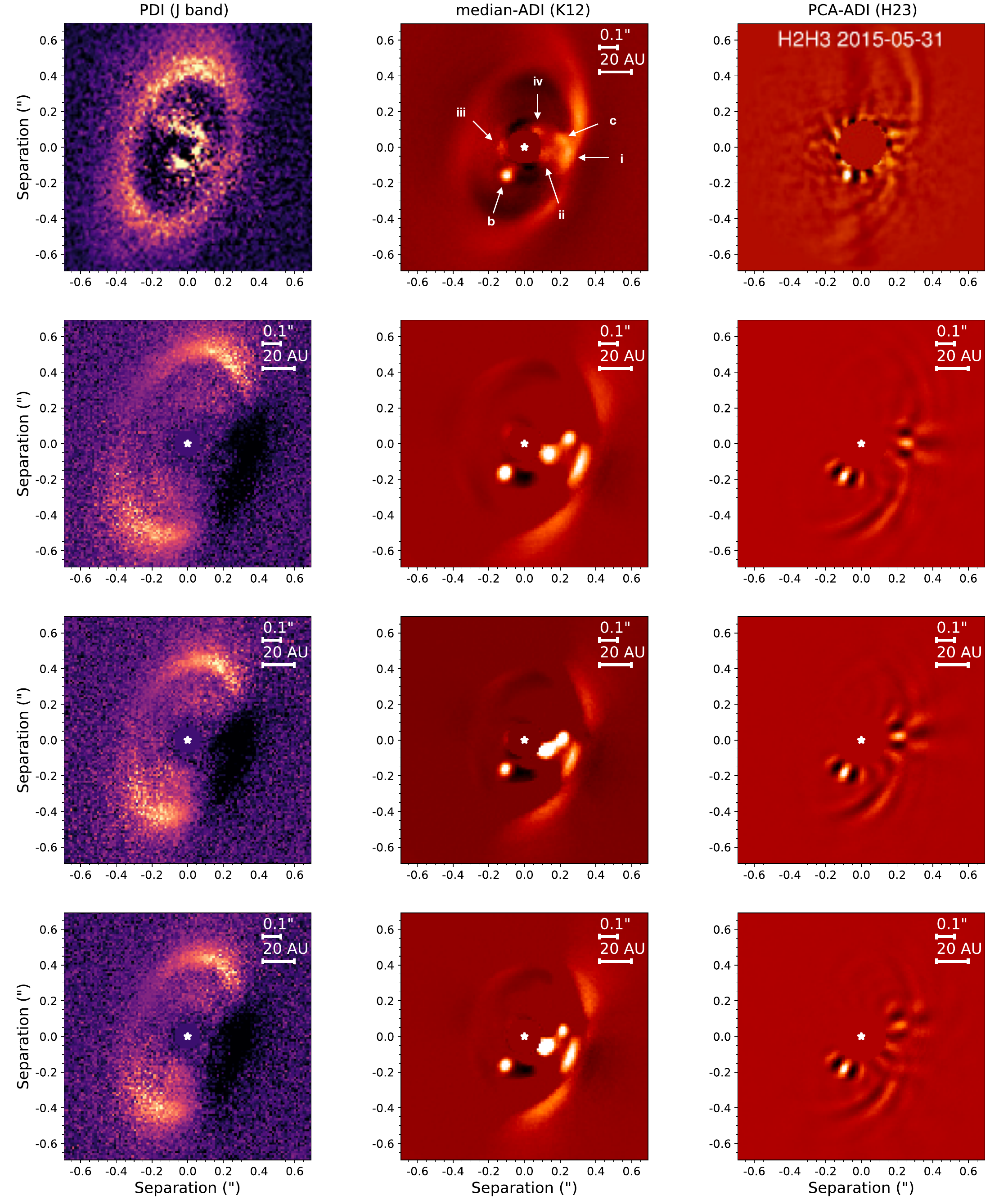}
 \caption{Near infrared observations of PDS 70 (top row) compared to the processed IR predictions obtained from simulations 1, 2 and 3 (second, third and last row, respectively). \textit{Left column:} Polarimetric Differential Imaging predictions in $J$ band (1.25 $\mu$m). \textit{Middle column:} average median-ADI images over the K1 and K2 bands (2.11/2.25 $\mu$m). \textit{Right column:} average PCA-ADI images over the H2 and H3 bands (1.59/1.67$\mu$m), obtained with 5 principal components subtracted.}
 \label{fig:DIeffects}
\end{figure*}

The primary focus of this work was to study the dynamics of the PDS 70 protoplanetary system. However, the close match between our radiative transfer predictions and reported observations in the infrared allows us to examine the biases in reported images of PDS 70 caused by observational and post-processing techniques.

Using a similar method to our post-processed K-band total intensity scattered light images (Section~\ref{IRpostprocessing}), we also predicted the map tracing the azimuthal component of polarised intensity ($Q_{\phi}$) in $J$ band (1.25 $\mu$m) for comparison to the coronagraphic IRDIS observations presented in \citet{Keppler2018}. 
We used the Q and U Stokes polarisation maps returned by \textsc{mcfost}, and applied steps \textit{ii} to \textit{v} detailed in Section~\ref{IRpostprocessing}: convolution with the observed PSF ($\sim$ 52-mas full width at half maximum), transmission of the apodised Lyot coronagraph, pixel scale resampling, and injection of shot noise at a similar level as the observations.
We then combined the Q and U images to obtain the $Q_{\phi}$ map, as is now standard for polarimetric differential imaging observations \citep[PDI; e.g.][]{Kuhn2001,Avenhaus2017}:
\begin{equation}
    Q_{\phi} = Q \cos(2\phi) + U \sin(2\phi),
\end{equation}
with
\begin{equation}
    \phi = \arctan\frac{x-x_0}{y-y_0}.
\end{equation}

Figure~\ref{fig:DIeffects} (left column) shows the predicted $Q_{\phi}$ maps for simulations 1 to 3. The size of the IR cavity is well reproduced in simulations 2 and 3.  
However we note that (1) the peak signal of the cavity edge lies along the PA of semi-major axis of the disc instead of a PA$\sim$0 deg and (2) a darker  than observed signature appears along the forward-scattered side of the disc, which both point towards a mismatch between our assumed grain distribution and the real one. Since the phase function of the degree of polarisation depends on both the grain size and the observed wavelength \citep[see e.g.][]{Murakawa2010}, it is beyond the scope of this work to search for the grain size distribution that reproduces the disc observations at all wavelengths.
Another minor difference between prediction and observation is the presence of an intra-cavity ring-like emission that is not observed in PDI. 
We hypothesise that a better match would be obtained if the simulations were run for more orbits of the companions as this would further deplete the cavity of small grains. In a similar way, this would likely reduce the intensity of feature \textit{ii} in the median-ADI images, and make it more in line with the fainter signal (with respect to the planets) seen in the observation.

We have also processed our infrared images using the smart principal component analysis (sPCA) algorithm implemented in \textsc{vip} \citep{Amara2012,Absil2013, GomezGonzalez2017}, similar to \citet{Keppler2018}. This is shown in the right column of Figure~\ref{fig:DIeffects}. We followed the exact same steps as described in Section~\ref{IRpostprocessing}, apart from step \textit{viii} where we used sPCA-ADI instead of median-ADI. Compared to the median-ADI images (centre column), the bulge in the middle of the forward-scattered edge of the cavity (feature \textit{i}) is attenuated, but several parallel arcs appear along the SW and NW parts of the edge. The number and intensity of these arcs increase with increasing number of principal components. We find the closest match to the observed image for 4 to 10 principal components subtracted. While increasing the number of principal components induces more geometric biases to extended disc signal, it nonetheless appears to help isolating the signal from planet c lying along the edge. Indeed, the blob gets flanked by characteristic negative ADI side lobes in azimuth, of increasing intensity with increasing number of principal components subtracted.

%If you want to present additional material which would interrupt the flow of the main paper,
%it can be placed in an Appendix which appears after the list of references.

%%%%%%%%%%%%%%%%%%%%%%%%%%%%%%%%%%%%%%%%%%%%%%%%%%

% Don't change these lines
\bsp	% typesetting comment
\label{lastpage}
\end{document}